\documentstyle[twocolumn,eqsecnum,aps,epsf,epsfig]{revtex}
\newcommand{\be}{\begin{equation}}
\newcommand{\ee}{\end{equation}}
\newcommand{\bea}{\begin{eqnarray}}
\newcommand{\eea}{\end{eqnarray}}

\newcommand{\lton}{\mathrel{\lower.9ex
                  \hbox{$\stackrel{\displaystyle <}{\sim}$}}}

\begin{document}
\title{First-Order Chiral Phase Transition in High-Energy Collisions:
Can Nucleation Prevent Spinodal Decomposition?}
\author{O.\ Scavenius$^{a,b}$, A.\ Dumitru$^c$, E.\ S.\ Fraga$^d$, J.\ T.\ Lenaghan$^{b,e}$, A.\ D.\ Jackson$^b$}
\address{
$^a$ NORDITA, Blegdamsvel 17, DK-2100 Copenhagen {\O}, Denmark\\
$^b$ The Niels Bohr Institute, Blegdamsvej 17, 
DK-2100 Copenhagen {\O}, Denmark\\
$^c$ Department of Physics, 
Columbia University, New York, NY 10027, USA \\
$^d$ Department of Physics, Brookhaven National Laboratory, 
Upton, NY 11973-5000, USA \\
$^e$ Department of Physics, Yale University, New Haven, Connecticut 06520, USA}
\maketitle   


\begin{abstract}
We discuss homogeneous nucleation in a first-order chiral phase transition 
within an effective field theory approach to low-energy QCD.  Exact decay 
rates and bubble profiles are obtained numerically and compared to analytic 
results obtained with the thin-wall approximation.  The thin-wall 
approximation overestimates the nucleation rate for any degree of 
supercooling.  The time scale for critical thermal fluctuations is 
calculated and compared to typical expansion times for high-energy 
hadronic or heavy-ion collisions. We find that significant supercooling 
is possible, and the relevant mechanism for phase conversion might be that of
spinodal decomposition. Some potential experimental signatures of 
supercooling, such as an increase in the correlation length of the scalar 
condensate, are also discussed.
\end{abstract}

\narrowtext      


\section{Introduction}

It is hoped that experiments in the near future at BNL-RHIC 
and CERN-LHC will discover evidence for the so-called quark-gluon plasma 
(QGP) in high-energy collisions of heavy-ions or protons~\cite{QGP}.
The QGP is the high-density primordial state of strongly interacting
matter that presumably existed in the early universe a few microseconds 
after the Big Bang~\cite{qgpBB,Ignatius:1994fr}.  In the QGP, chiral symmetry
is restored due to finite-temperature contributions to the effective
potential~\cite{Dolan:1974qd}  (This statement is only approximate since 
finite current-quark masses break chiral symmetry explicitly.  See below.).

Given our lack of detailed quantitative knowledge regarding low energy QCD, 
it is difficult to identify unambiguous observables that would, if found, 
demonstrate beyond reasonable doubt that the QGP had been produced.  Central 
ultra-relativistic collisions of heavy ions at RHIC and LHC will produce 
multiplicities on the order of $10^4$ hadrons per event~\cite{Back:2000gw}
 and raise the 
possibility of analyzing data event-by-event~\cite{Stock:1994ve}.  
Event-by-event fluctuations might provide an opportunity to discover new 
physics that is usually washed out by standard event averaging. In particular, 
if the phase transition is first-order with fast expansion so that strong 
supercooling or spinodal decomposition is possible, fluctuations in the 
rapidity density or in the transverse momentum spectra of hadrons and isospin 
fluctuations produced by the formation of domains of a coherent chiral 
condensate~\cite{rapi,bja,Scavenius:1999zc} could become interesting observables.
 
The nature of the QCD phase diagram in the temperature, $T$, and baryon 
chemical potential, $\mu$, plane has been intensively studied in recent 
years~\cite{Stephanov:1999zu,ove_dirk}.  Random matrix and effective model
calculations suggest that, at low $\mu$ and non-zero $T$, a smooth 
cross-over transition is expected and that, at low $T$ and non-zero $\mu$, 
the chiral phase transition should be of first order. Thus, a second-order 
critical point must exist in between these two limits.  The arguments 
suggesting this topology for the phase diagram are in no sense complete, and the 
phase transition could also be first-order for $\mu=0$.

In a first-order phase transition, the thermodynamic potential exhibits a 
metastable minimum corresponding to the symmetry restored phase for 
temperatures slightly less than the critical temperature, $T_c$.  This 
minimum gradually disappears as the system cools  and ends at a 
point of inflection, i.e.\ a ``spinodal instability''.  In a slowly expanding 
system, the phase transition would proceed through the nucleation of bubbles 
of the ``true vacuum'' state via thermal activation~\cite{Ignatius:1994fr,Enqvist:1992xw,Gleiser:1991rf,langer,linde,kapusta,ESF}.  In ultrarelativistic 
collisions of heavy ions or hadrons, it is clear that expansion is very fast 
compared to that of the early universe at the cosmological QCD transition 
(by a factor of $\sim 10^{18}$: $H\simeq 10^{23}$~sec$^{-1}$ versus $H \simeq 
10^{5}$~sec$^{-1}$).  Thus, it is necessary to investigate the time scale for 
thermal nucleation relative to that for expansion to see if the metastable 
chirally symmetric state of matter can reach the spinodal instability before 
nucleation is completed.

In this paper we shall consider a first-order chiral phase transition in a 
rapidly expanding background.  We take a linear $\sigma$-model coupled 
to quarks as our effective theory.  Treating the quarks as a heat 
bath which generates an effective potential for the soft modes of the chiral 
order parameter, we obtain an effective potential at the one-loop level.  
Somewhat below the critical temperature, this slightly asymmetric effective 
potential has three local minima.  Starting with the order parameter 
``localized'' in the metastable chirally symmetric state of matter, we then 
study critical bubble profiles and calculate the nucleation rate to the 
broken symmetry phase.  This will provide the temporal and spatial 
scales of thermal field fluctuations into the broken symmetry state.  We 
will consider both analytical and numerical computations.  Our numerical 
results represent ``exact'' calculations.  In the analytical treatment, we 
must turn to the thin-wall approximation in order to obtain closed results.  
The comparison between these two approaches will allow us to establish limits 
on the reliability of the thin-wall approximation.

This paper is organized as follows. Section II presents our effective 
theory.  Section III describes homogeneous nucleation and the methods 
used for both numerical and analytic computations of bubble profiles and 
nucleation rates.  Our results are discussed in Section 
IV.  Section V presents our conclusions and suggested directions for future 
investigation. We employ natural units, $\hbar=c=k_B=1$, throughout.


\section{Effective theory}

As an effective theory of the chiral symmetry breaking
dynamics, we assume the linear $\sigma$-model coupled
to quarks~\cite{Scavenius:1999zc,Gell-Mann:1960np,CsM95}:
\bea
{\cal L} &=&
 \overline{q}[i\gamma ^{\mu}\partial _{\mu}-g(\sigma +i\gamma _{5}
 \vec{\tau} \cdot \vec{\pi} )]q\nonumber\\
&+& \frac{1}{2}(\partial _{\mu}\sigma \partial ^{\mu}\sigma + \partial _{\mu}
\vec{\pi} \partial ^{\mu}\vec{\pi} )
- U(\sigma ,\vec{\pi})\quad.
\label{sigma}
\eea
The potential, which exhibits both spontaneously and explicitly broken 
chiral symmetry, is
\begin{equation} \label{T=0_potential}
U(\sigma ,\vec{\pi} )=\frac{\lambda ^{2}}{4}(\sigma ^{2}+\vec{\pi} ^{2} -
{\it v}^{2})^{2}-h_q\sigma\quad.
\end{equation}
Here $q$ is the constituent-quark field $q=(u,d)$. The
scalar field $\sigma$ and the pseudoscalar field $\vec{\pi} 
=(\pi_{1},\pi_{2},\pi_{3})$ together form a chiral field $\Phi = 
(\sigma,\vec{\pi})$. The parameters of the Lagrangian are chosen such that 
chiral $SU_{L}(2) \otimes SU_{R}(2)$ symmetry is spontaneously broken in the 
vacuum.  The vacuum expectation values of the condensates are $\langle\sigma\rangle 
={\it f}_{\pi}$ and $\langle\vec{\pi}\rangle =0$, where ${\it f}_{\pi}=93$~MeV 
is the pion decay constant. The explicit symmetry breaking term is due to the 
finite current-quark masses and is determined by the PCAC relation which gives 
$h_q=f_{\pi}m_{\pi}^{2}$, where $m_{\pi}=138$~MeV is the pion mass.  
This leads to $v^{2}=f^{2}_{\pi}-{m^{2}_{\pi}}/{\lambda ^{2}}$.  The value of 
$\lambda^2 = 20$ leads to a $\sigma$-mass, 
$m^2_\sigma=2 \lambda^{2}f^{2}_{\pi}+m^{2}_{\pi}$, equal to 600~MeV.  In 
mean field theory, the purely bosonic part of this Lagrangian exhibits a 
second-order phase transition~\cite{Pisarski:1984ms} at $T_c=\sqrt{2}v$ if 
the explicit symmetry breaking term, $h_q$, is dropped.  For $h_q\ne 0$, the 
transition becomes a smooth crossover from the restored to broken symmetry 
phases.  For $g>0$, the finite-temperature one-loop effective potential also 
includes the following contribution from the quarks:
\be\label{T>0_potential}
V_q(\Phi) = - d_q T
\int \frac{d^3k}{(2\pi)^3} \log\left(1+e^{-E/T}\right)\quad.
\ee
Here, $d_q=24$ denotes the color-spin-isospin-baryon charge degeneracy of the 
quarks. We note that $V_q(\Phi)$ depends on the order parameter $\Phi$ through 
the effective mass of the quarks, $m_q=g\sqrt{\Phi^2}$, which enters into 
the expression for the energy, $E = \sqrt{k^2 + g^2 \Phi^2}$.  In our 
phenomenological approach, we will assume that the quarks constitute the 
heat bath in which the long-wavelength modes of the chiral field, i.e.\ the 
order parameter, evolve. Thus, we will take the effective potential to be
\be \label{eq:pot1}
V_{\rm eff} \equiv U+V_q\quad.
\ee

To arrive at this form for the effective potential we
consider a system of quarks and antiquarks in thermodynamical
equilibrium at temperature $T$ (in~\cite{ove_dirk}
$V_{\rm eff}$ is derived also at finite baryon-chemical potential;
here we shall consider only the case of baryon-symmetric matter).
The partition function reads
\begin{equation}
{\cal Z}=
\int{\cal D}\bar{q}\, {\cal D}q\, {\cal D}\sigma\, {\cal D}\vec{\pi}\;
\exp\left[ - \int_0^{1/T} dt \int_{\cal V} d^3 {\vec {x}}\, {\cal L} \right]
\,\, .
\end{equation}
${\cal V}$ is the volume of the system.
In mean-field approximation the chiral fields in the
Lagrangian are replaced by their expectation values, which we will
denote by $\sigma$ and $\vec{\pi}$ in the following.
Then, up to an overall normalization factor,
\begin{eqnarray}
{\cal Z} & = & {\cal N}_U
\int{\cal D}\bar{q}\, {\cal D}q\; 
\exp\Biggl\{  - \int_0^{1/T} dt \int_{\cal V} d^3 {\vec {x}} \nonumber\\
& &
 \bar{q} \left[ i\gamma ^{\mu}\partial _{\mu}-
g(\sigma +i\gamma _{5}\vec{\tau} \cdot \vec{\pi})\right]q
 \Biggr\} \nonumber \\
& = & {\cal N}_U \,
{\rm det}_p \Bigl\{ \bigl[ p_{\mu}\gamma^{\mu} -
g(\sigma +i\gamma_{5}\vec{\tau} \cdot \vec{\pi}) \bigr]/T\Bigr\}\,\, ,
\end{eqnarray}
where
\be
{\cal N}_U = \exp\left(-\frac{{\cal V}U(\sigma ,\vec{\pi} )}{T}\right)~.
\ee
Taking the logarithm of ${\cal Z}$, the determinant of the Dirac
operator can be evaluated in the standard fashion~\cite{kapustaFFT},
and we finally obtain the grand canonical potential
\begin{equation}
\Omega (T)=-\frac{T}{{\cal V}}\log {\cal Z} =
U + \tilde{V}_q(T)~,
\label{ptotsig}
\end{equation}
where
\begin{equation}
\tilde{V}_{q}(T) = -d_{q} \int \frac{d^3{\vec{p}}}{(2\pi)^3}
\left\{E+T\, \log \left[1+e^{{-E}/{T}} \right]
 \right\}\,\, .
\label{pqq}
\end{equation}
For our purposes, the zero-temperature contribution to $\tilde{V}_{q}$
can be partly absorbed into $U$ via a standard renormalization of the bare
parameters $\lambda^2$ and $v^2$. However, a logarithmic
correction from the renormalization scale
remains and is neclected in the following (see \cite{ove_dirk}
for more details). The finite-$T$ part defines
$V_q$ as given in eq.~(\ref{T>0_potential}).

We now turn to a discussion of the shape of the effective
potential at various temperatures.
\begin{figure}[htbp]
\centerline{\hbox{\epsfig{figure=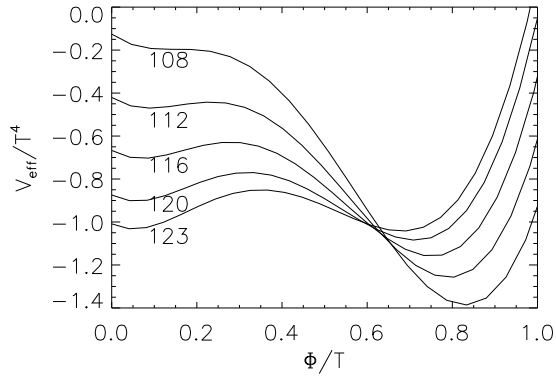,width=8cm}}}
\caption{The one-loop finite-$T$ effective potential
for $T_{\rm sp} \le T \le T_c$ and coupling constants $g=5.5$,
$\lambda^2=20$. 
The curves are labeled by the temperature in MeV.}
\label{potfig}
\end{figure} 
For sufficiently small $g$ one still finds the above-mentioned smooth 
transition between the two phases.  At larger coupling to the quarks, however,
the effective potential exhibits a first-order phase transition.  
For temperatures near the critical temperature, $V_{\rm eff}$ displays 
a local minimum $\Phi = \Phi_f(T) \simeq 0$ which is separated by a barrier 
from another local minimum at $\Phi=\Phi_t(T)>0$.  (There is another local 
minimum for negative $\Phi$ which is of higher energy and need not concern us.)
These two minima are degenerate at $T=T_c$. For example,
$g=5.5$ leads to a critical temperature of 
$T_c=123.7$~MeV. Unless stated differently, throughout the manuscript
we shall adopt the values
$g=5.5$ for the quark-field coupling and $\lambda^2=20$ for the self-coupling
of the chiral fields.

Chiral symmetry is (approximately) restored for $T > T_c$.  
The minimum at $\Phi=\Phi_t$ becomes the absolute minimum as the temperature 
is reduced below $T_c$.  As the temperature is lowered, the local minimum 
at $\Phi \approx 0$ approaches the intervening maximum.  These two 
extrema meet and form an inflection point at the spinodal temperature, 
$T_{\rm sp}$.  Only the broken symmetry minimum remains for $T < T_{\rm sp}$.  
For our standard
choice of parameters, $T_{\rm sp} = 108$~MeV.  The potential in 
the $\sigma$-direction is shown in Fig.~\ref{potfig} for the range of 
interest, $T_{\rm sp} \le T \le T_c$; see also 
Ref.~\cite{Scavenius:1999zc}\footnote{The right diagram in Fig.\ 1 of 
Ref.~\cite{Scavenius:1999zc} is incorrectly indicated as corresponding 
to $T=100$~MeV.  It actually represents $T=108$~MeV.}.  Note the 
rather small barrier between the two local minima of $V_{\rm eff}$ at 
$T_c$.  The first-order transition for $g=5.5$ is weak compared to those 
conventional in bag model calculations~\cite{kapusta}.  This weakness is 
due to the fact that gluons, absent in the present linear $\sigma$-model, 
are treated as free fields (for $T \ge T_c$) in the bag model.  It is not 
our goal here to argue which approximation is more reasonable.  Rather, we 
view our approach as complementary to previous studies which employed the bag
model~\cite{Ignatius:1994fr,Enqvist:1992xw,kapusta}.  The existence of a 
local minimum in $V_{\rm eff}$ raises the possibility of super-cooling 
effects since the true broken symmetry minimum can only be reached by 
tunnelling.  The weakness of our first-order transition is likely to 
overestimate tunnelling rates and underestimate the effects of supercooling.  

\begin{figure}[htp]
\centerline{\hbox{\epsfig{figure=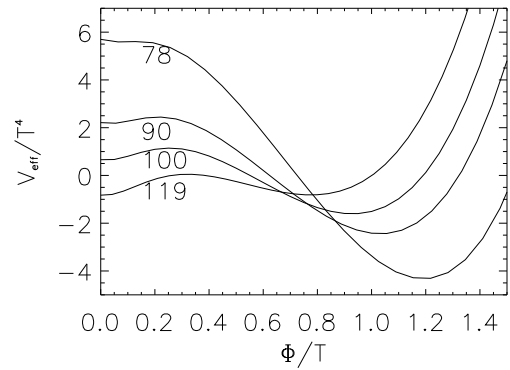,width=8cm}}}
\caption{The one-loop finite-$T$ effective potential
for $T_{\rm sp} \le T \le T_c$ and coupling constants $g=10$,
$\lambda^2=20$. The curves are labeled by the temperature in MeV.}
\label{potfig_g10}
\end{figure} 
In fact, increasing the coupling to the quarks leads to a larger barrier
between the two local minima of the effective potential, as seen in
Fig.~\ref{potfig_g10}. In other words, the first-order transition
becomes stronger,
with larger latent heat. Furthermore, the spinodal temperature $T_{\rm sp}$
decreases considerably, to about $\simeq78$~MeV. As will become clear below,
both effects (the higher barrier and the lower temperatures) act to reduce
the decay rate of the metastable minimum as compared to our
``standard'' choice of couplings corresponding to the potential shown
in Fig.~\ref{potfig}.

\begin{figure}[htp]
\centerline{\hbox{\epsfig{figure=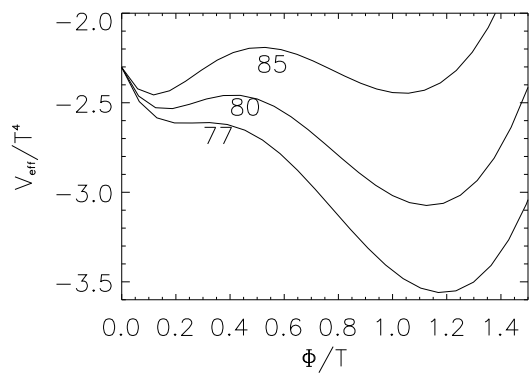,width=8cm}}}
\caption{The one-loop finite-$T$ effective potential
for $T_{\rm sp} \le T \le T_c$ and coupling constants $g=5.5$,
$\lambda^2=2.2$. The curves are labeled by the temperature in MeV.}
\label{potfig_lam2}
\end{figure} 
There is also the possibility that the $\sigma$-meson is much lighter
at $T_c$ than at $T=0$~\cite{GGP} (i.e.\ $\lambda^2$ is small) if QCD is
near a chiral critical point.
Chosing for example $\lambda^2\simeq2.2$ with the pion decay constant
$f_\pi$ and vacuum mass $m_\pi$ fixed, such that $v^2=0$, leads to the
potential shown in Fig.~\ref{potfig_lam2}. Clearly, decreasing $\lambda^2$
in this theory reduces $T_c$ very much, to less than $100$~MeV.
Reducing $g$ has practically no effect, except that the first-order phase
transition disappears at $g\lton4$. Such low $T_c$ appear to be excluded
by present lattice QCD results~\cite{Laermann:1996xn}, and will
therefore not be considered further.

In order to obtain approximate analytic formulas, it is convenient to express 
$V_{\rm eff}$ over the range $0 \le \Phi \le T$ in the familiar 
Landau-Ginzburg form (see also~\cite{Enqvist:1992xw})
\be \label{LandGinz}
V_{\rm eff} \approx \sum_{n=0}^4 a_n \, \Phi^n\quad.
\ee
This form is adopted purely for reasons of convenience and no deeper meaning 
should be attached to it.  It is, for example, obviously incapable of 
reproducing all three minima of $V_{\rm eff}$.  We shall use this expansion 
both for a limited range of temperatures ($T_{\rm sp} \le T \le T_c$) and 
for a limited range of field strengths ($0 \le \Phi \le T$) which includes 
the two local minima of interest.  With these restrictions, this polynomial 
form is found to provide a quantitative description of $V_{\rm eff}$.  The 
coefficients $a_n$ are determined for each $T$ by performing a least squares 
fit to the true effective potential.


\section{Homogeneous Nucleation}
We wish to consider the physical situation (realizable for $T_{\rm sp} \le 
T \le T_c$) in which a scalar field, corresponding to the $\sigma$-direction, 
is initially localized in the metastable restored symmetry minimum of an 
asymmetric double-well potential.  (We neglect the higher-energy local 
minimum with $\Phi < 0$.)  It is well known that such a system can decay 
by the nucleation of thermally activated bubbles of the true vacuum inside 
the false one~\cite{Ignatius:1994fr,Enqvist:1992xw,Gleiser:1991rf,langer,linde,kapusta,ESF}. The nucleation rate per unit volume per unit time is 
expressed as 
\be \label{decrate}
\Gamma={\cal P}~ e^{-F_b/T}\quad,
\ee
where $F_b$ is the free energy of a critical bubble and where the
prefactor ${\cal P}$ provides a measure of the saddle point of the
Euclidean action in functional space.  It is conventional to write 
${\cal P}$ as a product of the bubble's growth rate and a factor proportional 
to the ratio of the determinant of the fluctuation operator around the bubble 
configuration to that around the homogeneous metastable state.  The  
following analysis will concentrate on the exponential barrier penetration 
factor, and we will approximate ${\cal P}$ by $T^4$.  This approximation is 
known to give an overestimate of $\Gamma$, see e.g.~\cite{linde,kapusta}.
The reason for this focus is that $F_b$ is necessarily infinite at 
$T_c$ and zero at $T_{\rm sp}$.  The exponential factor is thus expected 
to make the dominant contribution to the structure of $\Gamma$.

In order to determine the role of bubble nucleation in the evolving system 
and to be able to compute the decay rate $\Gamma$, it is necessary to 
study the critical bubble and some of its features.  The critical 
bubble is a radially symmetric, static solution of the Euler-Lagrange field 
equations that satisfies the boundary condition $\Phi (r \to \infty) 
\longrightarrow \Phi_f$ where $\Phi_f$ is the false vacuum of the effective 
potential.  Energetically, this boundary condition corresponds to an 
exact balance between volume and surface contributions which defines the 
critical radius, $R=R_c$.  The critical bubble is unstable with respect to 
small changes of its radius.  For $R < R_c$, the surface energy 
dominates, and the bubble shrinks into the false vacuum.  For $R>R_c$, the 
volume energy dominates, and the bubble grows driving the decay process.

The critical bubble can be found by minimizing the free energy
\be
F_b(\Phi,T)=4\pi\int \, r^2\, {\rm d}r\, \left[\frac{1}{2}\left(\frac{{\rm d}
\Phi}{{\rm d}r}\right)^2
+V_{\rm eff}(\Phi,T)\right]\quad,
\label{free}
\ee
with respect to the field $\Phi$.  This leads to 
\be
\frac{{\rm d}^2\Phi}{{\rm d}r^2}+\frac{2}{r}\frac{{\rm d}\Phi}{{\rm d}r}-
\frac{\delta V_{\rm eff}(\Phi,T)}{\delta\Phi}=0\quad,
\label{bounce}
\ee
where
\bea
\frac{\delta V_{\rm eff}(\Phi,T)}{\delta\Phi} &=& \lambda(\Phi^2-v^2)\Phi-h_q
\nonumber \\ 
&+& \frac{g^2\Phi d_q}{\pi^2} \int
{\rm d}p\frac{p^2}{E}\frac{1}{\exp(E/T)+1}\quad.
\eea
(It has been assumed in Eq.\,(\ref{free}) that the effective potential is 
shifted so that $V_{\rm eff} (\Phi_f , T) = 0$.)  The desired solution of 
Eq.~(\ref{bounce}) is found by imposing the boundary condition $\Phi (r \to 
\infty) \longrightarrow \Phi_f$.  

For numerical purposes it is more convenient to impose boundary conditions 
at $r = 0$ by setting $\Phi' (0) = 0$ and $\Phi (0)$ equal to an ``escape 
value'' $\Phi_e$.  It is straightforward to integrate Eq.~(\ref{bounce}) to 
obtain the bubble profile $\Phi(r)$ for a given temperature.  The escape 
value is then adjusted until the desired boundary conditions at $r \to \infty$ 
are realized.  This process can be facilitated by smoothly joining the 
numerical values of $\Phi (r)$ to the analytic solution to Eq.\,(\ref{bounce}) 
that is regular as $r \to \infty$.  Close to the critical temperature, $\Phi 
(r)$ has a simple qualitative behaviour.  It stays close the escape value 
$\Phi_e$ for $r < R_c$ and then makes a rapid transition to the false 
vacuum, $\Phi_f$.  For obvious reasons, $R_c$ is identified as the radius of 
the critical bubble.  Once the profile of the critical bubble has been 
found, the free energy $F_b$ is obtained by computing the integral in 
Eq.~(\ref{free}) numerically.  Finally, the decay rate can be calculated 
using the approximate prefactor ${\cal P} = T^4$.  

A simple physical analogue of this procedure can be obtained by turning 
the potentials in Fig.~\ref{potfig} upside down and considering the mechanical 
motion of a point particle moving in the inverted potential.  The 
problem of determining the critical bubble profile is then seen to be 
equivalent to finding that point on the true vacuum hill such that a ball 
starting at rest and obeying the equation of motion, Eq.~(\ref{bounce}), would 
descend and land at rest on the top of the false vacuum hill at the point 
$\Phi_f$.  The starting point is evidently the escape point.  It does not 
correspond to the true vacuum point but rather to some smaller value.  This 
mechanical analogy also makes it clear that $-V_{\rm eff} (\Phi_e )$ must be greater 
than $-V_{\rm eff} ( \Phi_f )$ due to the presence of the second, 
``frictional''  term in Eq.~(\ref{bounce}).  As we shall see, the critical 
radius $R_c$ diverges as $T \to T_c$.  The frictional term then becomes small 
in some sense.  The various approximate methods for dealing with this 
frictional term approximately are all described as ``the thin-wall 
approximation''.  

Although the thin-wall approximation is not expected to be valid near the 
spinodal, it should provide a reliable description of nucleation near $T_c$.  
Given the frequency of its adoption and its simplicity, we find it useful 
to apply the thin-wall approximation to the present problem.  A quartic 
potential such as Eq.\,(\ref{LandGinz}) can always be rewritten in the form
\be
W(\varphi)=\alpha ~ (\varphi^2-a^2)^2+j\varphi\quad.
\label{www}
\ee
Specifically, we have
\bea
\alpha &=& a_4\quad, \\
a^2 &=& \frac{1}{2}\left[ -\frac{a_2}{a_4}
+\frac{3}{8}\left(\frac{a_3}{a_4}\right)^2  \right]\quad, \\
j &=& a_4\left[ \frac{a_1}{a_4}-
\frac{1}{2}\frac{a_2}{a_4}\frac{a_3}{a_4}+
\frac{1}{8}\left(\frac{a_3}{a_4}\right)^3   \right]\quad, \\
\varphi&=&\Phi + \frac{1}{4}\frac{a_3}{a_4} \ . 
\eea
The new potential $W(\varphi)$ reproduces the original $V_{\rm eff} ( \Phi )$ 
up to a shift in the zero of energy.  We are interested in the effective 
potential only between $T_c$ and $T_{\rm sp}$.  At $T_c$, we will have two 
distinct minima of equal depth.  This clearly corresponds to the choice $j = 0$
in Eq.\,(\ref{www}) so that $W$ has minima at $\varphi = \pm a$ and a maximum 
at $\varphi = 0$.  The minimum at $\varphi = -a$ and the maximum move closer 
together as the temperature is lowered and merge at $T_{\rm sp}$.  Thus, 
the spinodal requires $j/\alpha a^3 = -8/3\sqrt{3}$ in Eq.\,(\ref{www}). 
In the present case, the parameter $\alpha$ is essentially independent of 
temperature, and $a$ rises roughly linearly by some 3\% as the temperature 
falls from $T_c$ to $T_{\rm sp}$.  The parameter $j/\alpha a^3$ falls roughly 
linearly from $0$ (at $T_c$) to $-8/3\sqrt{3}$ (at $T_{\rm sp}$).

The explicit form of the critical bubble in 
the thin-wall limit is then given by~\cite{ESF}
\be \label{bubproftw}
\varphi_c (r;\xi,R_c)=\varphi_f + \frac{1}{\xi\sqrt{2\alpha}}
\left[ 1-\tanh \left( \frac{r-R_c}{\xi} \right) \right]\quad,
\ee
where $\varphi_f$ is the new false vacuum, $R_c$ is the radius of the 
critical bubble, and $\xi=2/m$ with $m^2\equiv W''(\varphi_f)$ is a measure 
of the wall thickness.  The thin-wall limit corresponds to $\xi/R_c\ll 1$ 
\cite{ESF}, which can be rewritten as $(3|j|/8\alpha a^3)\ll 1$.  This small 
parameter has the value of $1/\sqrt{3}$ at the spinodal, which suggests 
that the thin-wall approximation might be qualitatively reliable for 
our purposes.  Nevertheless, it merits a quantitative check.  In terms of 
the parameters $\alpha$, $a$, and $j$ defined above, we find 
\bea
\varphi_{t,f} &\approx& \pm a - \frac{j}{8\alpha a^2}  \\
\xi &=& \left[ \frac{1}{\alpha (3\varphi_f^2-a^2)} \right]^{1/2}
\label{twcorlength}
\eea
in the thin-wall limit.  Determination of the critical radius requires 
the surface tension, $\Sigma$, defined as 
\be
\Sigma\equiv \int_0^{\infty}{\rm d}r~\left( \frac{{\rm d}
\varphi_b}{{\rm d}r} \right)^2 
\approx \frac{2}{3\alpha\xi^3} \ .
\ee
The critical radius then becomes 
\be
R_c = \frac{2\Sigma}{\Delta W}\quad, \label{EqRc}
\ee
where
\be
\Delta W \equiv V(\Phi_f)-V(\Phi_t) \approx 2 a | j | \quad.
\ee
The energy of a critical bubble is finally given by
\be
E_c=\frac{4\pi\Sigma}{3}R_c^2\quad.
\ee
From knowledge of $F_b = E_c$, one can evaluate the nucleation rate $\Gamma$.  
In calculating thin-wall properties, we shall use the approximate forms 
for $\phi_t$, $\phi_f$, $\Sigma$, and $\Delta W$ for all values of the 
potential parameters.


\section{Results}

\begin{figure}[htp]
\centerline{\hbox{\epsfig{figure=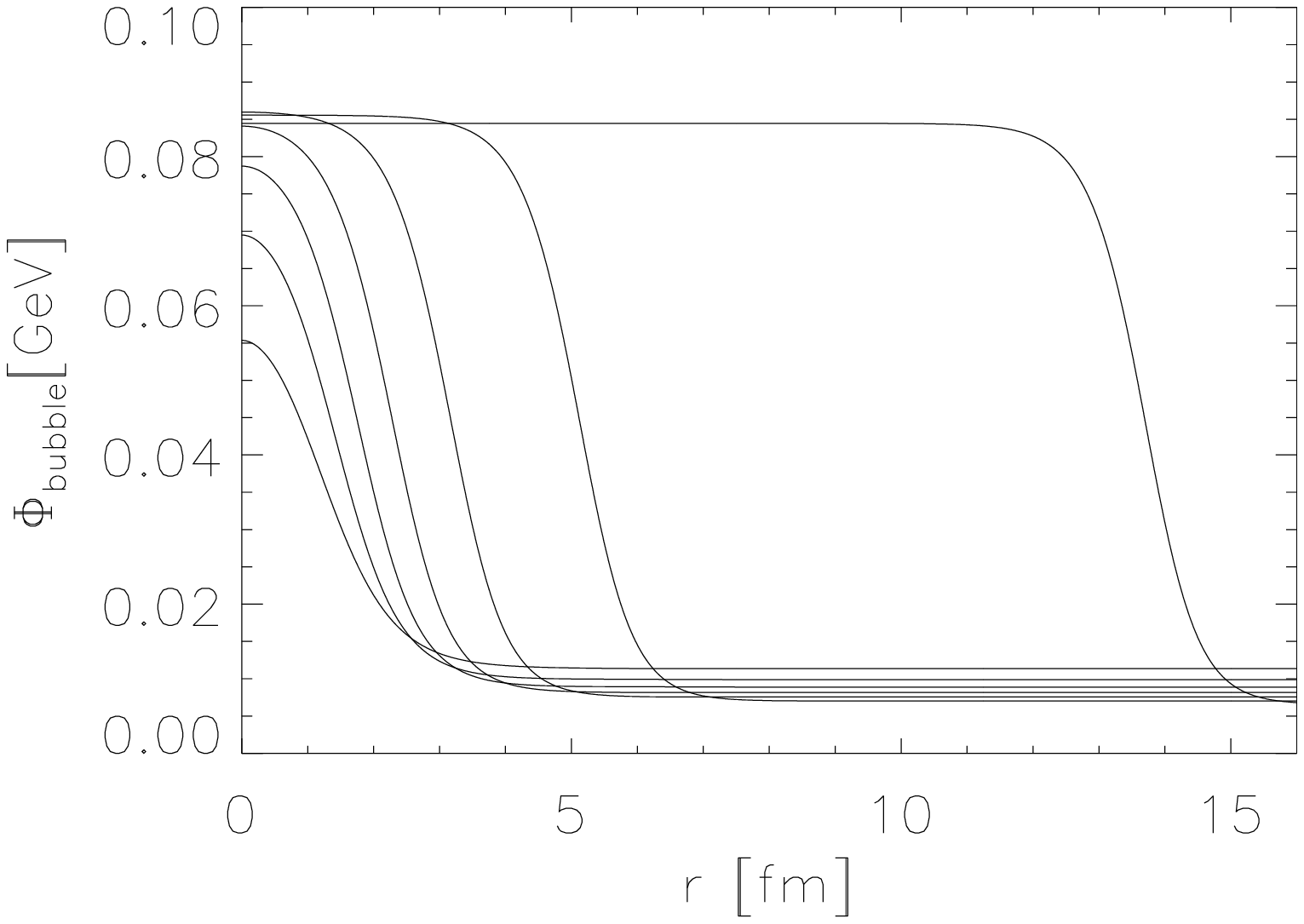,width=8cm}}}
\caption{Exact bubble profiles for temperatures 110, 112, 114, 116, 118, 120, 
and 122~MeV.  Bubble size increases monotonically with temperature.  Recall 
that $T_c = 123.7$~MeV and $T_{\rm sp} = 108$~MeV.}
\label{Figbubproe}
\end{figure}

\begin{figure}[htp]
\centerline{\hbox{\epsfig{figure=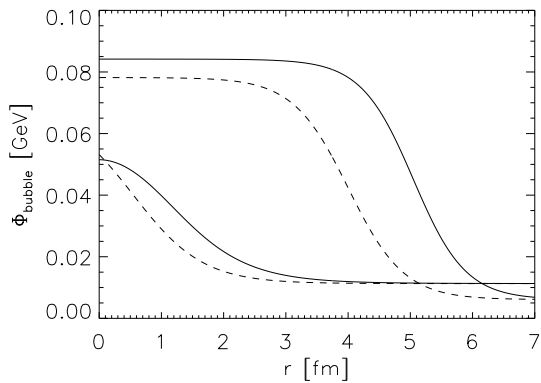,width=8cm}}}
\caption{Critical bubble profiles for temperatures 110 and 120~MeV.  The 
solid line indicates exact numerical results.  The dashed line shows the 
results obtained with the thin-wall approximation.}
\label{bubcomp}
\end{figure}

We begin by showing the critical bubble profiles obtained from the exact 
numerical solution of Eq.~(\ref{bounce}) in Fig.~\ref{Figbubproe}. 
For small supercooling the bubble profiles exhibit a pronounced ``core'' with 
$\Phi \simeq \Phi_t$ separated by a relatively thin wall from the region
$\Phi=\Phi_f$. On the other hand, critical bubbles become ``coreless'' for 
stronger supercooling; their thickness is of the same order as the radius, and
the field at the escape point, $\Phi(r=0)$, does not reach $\Phi_t$.  
For comparison, Fig.~\ref{bubcomp} shows bubble profiles for temperatures 
$T= 110$ and $120$~MeV as obtained from the numerical solution of 
Eq.~(\ref{bounce}) and from the thin-wall approximation.  In this case, 
we have used the same polynomial approximation to $V_{\rm eff}$ in both 
exact and thin-wall calculations so that discrepancies can be associated 
cleanly with the thin-wall approximation.  (The critical bubble profile is 
quite sensitive to the precise form of $V_{\rm eff}$.  This is particularly 
true when $T \approx T_c$ and the bubble radius is large.) 
As expected, the thin-wall approximation provides a qualitatively 
reasonable description of the critical bubble profile.    
\begin{figure}[htp]
\centerline{\hbox{\epsfig{figure=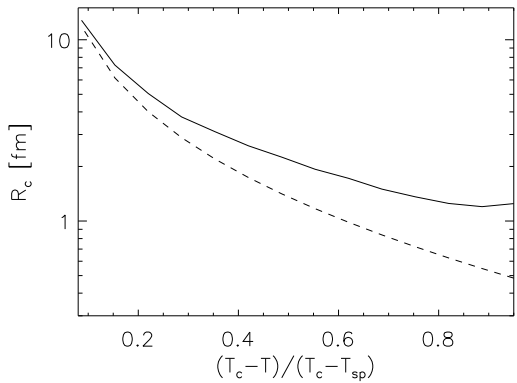,width=8cm}}}
\caption{Critical radii computed numerically (solid line) and with the
thin-wall approximation (dashed line) as functions of the degree of 
supercooling.}
\label{Figradii}
\end{figure} 
In Fig.~\ref{Figradii}, we compare the critical radii $R_c$ obtained from the
two calculations as a function of temperature. For the exact numerical
calculation, $R_c$ has been extracted from the bubble profile shown in
Fig.~\ref{Figbubproe} as the point at which the curvature of the profile 
changes sign \footnote{This definition of $R_c$ is somewhat arbitrary.  
This figure is intended only for the comparison of exact and thin-wall 
results.  Our numerical calculations do not employ $R_c$ in any of the 
quantities discussed below.}.  This figure makes it clear that the 
thin-wall approximation becomes quantitatively correct
as $T \to T_c$ and $R_c \to 
\infty$ as expected.  The fact that the thin-wall approximation 
systematically underestimates $R_c$ will be reflected in a systematic 
overestimate of the decay rate of the metastable state.

\begin{figure}[htp]
\centerline{\hbox{\epsfig{figure=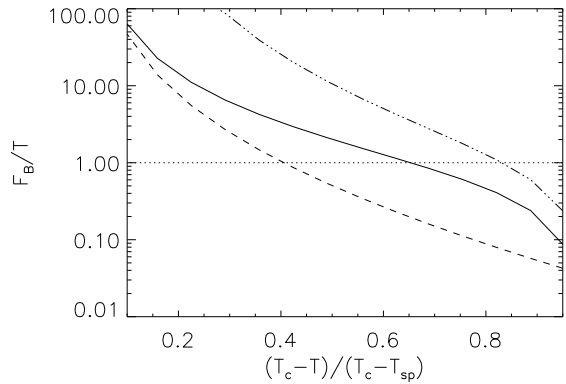,width=8cm}}}
\caption{The free energy of the critical bubbles as a function of the 
degree of supercooling.  The solid line shows the exact (numerical) result, 
the dashed line shows the thin-wall approximation, and the dotted line 
shows exact results for the modified effective potential discussed in 
the text.}
\label{FigFb}
\end{figure}

This expectation is also supported by Fig.~\ref{FigFb} which shows 
$F_b / T$ as a function of the degree of supercooling, $x$, defined as 
$x = (T_c - T)/(T_c - T_{\rm sp})$.  This figure also allows us to 
calibrate the temperature range of primary interest in the 
present work.  We have chosen to concentrate on the relatively violent 
behaviour of the exponential factor in Eq.\,(\ref{decrate}) and to 
treat the prefactor, ${\cal P}$, crudely.  Given, the dramatic changes 
in $F_b / T$ as a function of supercooling shown in this figure, this 
focus seems appropriate.  For temperatures such that $F_b / T > 1$, 
$\Gamma$ will be strongly suppressed by the exponential factor, 
and the system is likely to remain in the metastable state for times 
comparable to the expansion time of the system.  For lower temperatures, the 
exponential factor is unimportant, and $\Gamma$ is determined primarily 
by the prefactor, ${\cal P}$.  At such temperatures, it becomes necessary to 
make a more detailed comparison of the decay rate with the expansion rate of 
the system in order to decide the fate of the metastable state.  Such 
comparison requires a more reliable description of ${\cal P}$ than 
that adopted here.  The exact results shown in the figure suggest that the 
system is likely to supercool for some two-thirds of the way from $T_c$ to 
$T_{\rm sp}$.  By contrast, the thin-wall results yield the significantly 
smaller value of $x \approx 0.4$.  
Given the weakness of the first-order phase transition predicted by our 
model with $g = 5.5$, it is remarkable that the system can come so close 
to the spinodal temperature before the stability of the metastable state 
is seriously challenged.  It is to be expected that effective potentials 
which maintain the same $T_c$ and $T_{\rm sp}$ but which have a higher 
barrier between the phases at $T_c$ would permit the metastable state 
to come even closer to the spinodal temperature.      
\begin{figure}[htp]
\centerline{\hbox{\epsfig{figure=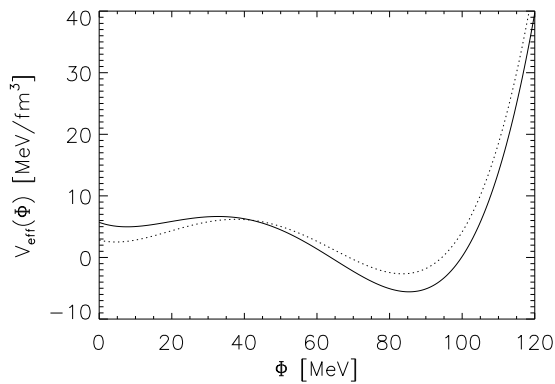,width=8cm}}}
\caption{The effective potential at $T=116$~MeV.  The solid line shows the 
original $V_{\rm eff}$ from Fig.~\ref{potfig}.
The dotted line shows the modified $V_{\rm eff}$ as discussed in the text.}
\label{potcom}
\end{figure} 
This point can be illustrated by considering an artificially modified 
effective potential with a higher barrier.  We construct this potential 
from our quartic approximations by arbitrarily multiplying the parameter 
$j$ at each temperature by a factor of $(T_c - T)/(T_c - T_{\rm sp})$.  
Evidently, this modification has no effect on $V_{\rm eff}$ at the 
critical temperature (where $j$ is $0$) or at the spinodal (where the 
additional factor is $1$).  The maximum difference in effective potentials 
lies halfway between these extremes and is illustrated in Fig.~\ref{potcom}.  
While the difference between these potentials is not dramatically large, 
it is sufficient to raise $F_b / T$ from the value of $2$ (for the original 
$V_{\rm eff}$) to $10$.  Following exponentiation, this leads to a reduction 
of $\Gamma$ by a factor of almost $3000$.  The values of $F_b / T$ for this 
modified potential, shown in Fig.~\ref{FigFb}, suggest stability of the 
metastable phase for all $x < 0.83$.  

It is now a simple matter to obtain $\Gamma$, the nucleation 
probability per unit time and volume, and to determine the space-time 
scales for homogeneous nucleation.  This decay rate must be compared with 
the expansion rate of the central region of a high-energy collision.  
Certain obvious conditions apply.  For example, the critical radius $R_c$
cannot exceed the horizon.  Otherwise, the center of the bubble cannot 
be causally connected to its surface, and the hydrostatic balance of surface 
and volume energy is thus impossible.  In addition, there is a time scale 
associated with a spontaneous coherent thermal fluctuation of energy $E_c$ 
with a length-scale of $R_c$.  If that time-scale is larger than the inverse 
expansion rate, bubble nucleation can be regarded as a slow process, and deep 
supercooling is to be expected.  (This is the condition that $F_b / T \gg 1$ 
discussed above.)  To our knowledge, these questions have not previously 
been addressed in studies of homogeneous nucleation in the QCD phase 
transition as found in hadronic collisions.

To simplify our considerations, we assume that the central rapidity region 
exhibits one-dimensional (longitudinal) Hubble flow~\cite{Bjorken:1983qr} 
with an expansion rate (i.e.\ Hubble constant) $H=1/t$, where $t$ denotes 
proper time.  Even in large nuclei, transverse expansion can increase the 
local expansion rate considerably when the system has cooled down to $T 
\approx T_c$~\cite{Dumitru:1999ud}.  Reasonable expansion rates are $H 
\approx 0.1-1$~fm$^{-1}$~\cite{Scavenius:1999zc,Dumitru:1999ud}. 
\begin{figure}[htp]
\centerline{\hbox{\epsfig{figure=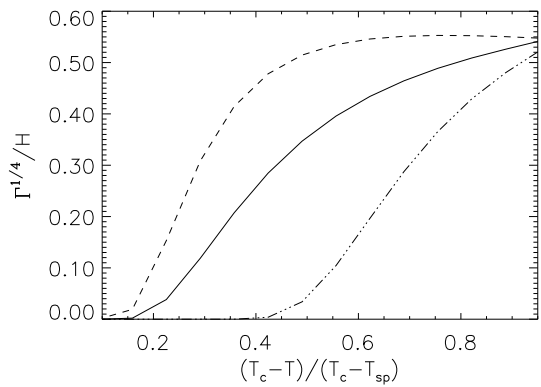,width=8cm}}}
\caption{$\Gamma^{1/4}$ divided by an assumed expansion rate of $H = 
1$~fm$^{-1}$ as a function of temperature.  The solid line shows the exact 
(numerical) result, the dashed line shows the thin-wall approximation, and 
the dotted line shows exact results for the modified potential.}
\label{FigRatio}
\end{figure} 
We have chosen $\Gamma^{1/4}/H$ as a measure for the comparison of nucleation 
and expansion~\cite{Gleiser:1991rf}.  The results shown in Fig.~\ref{FigRatio} 
assume an expansion of $H=1$~fm$^{-1}$.  When the nucleation rate is much 
larger than the expansion rate, it is reasonable to adopt standard 
thermodynamic techniques and to describe the phase transition by an idealized 
Maxwell construction~\cite{kapusta}.  Our results suggest that the Maxwell 
construction is not likely to provide a reliable description of the phase 
transition dynamics.  While the results of Fig.~\ref{FigRatio} do not provide 
an unambiguous prediction for the fate of the metastable state, they do make 
it appear likely that the rapidly expanding system has an appreciable 
probability of remaining in the restored symmetry phase even close to the 
spinodal instability. Thus, at least some fraction of all heavy ion events 
should show traces of this non-equilibrium transition.  This underscores the 
importance of the event-by-event analysis of data~\cite{Stock:1994ve}.

We now consider some potential consequences of a non-equilibrium transition 
associated with strong supercooling relative to an adiabatic transition.
First, the character of the phase transition will certainly
influence global observables and the expansion of the hot system.
For example, a non-equilibrium transition is always associated with
entropy production which can be as large as $30\%$~\cite{kapusta}. 
This would be reflected in the multiplicity of hadrons in the final state.
Moreover, phase coexistence in equilibrium leads to a distinct hydrodynamic 
expansion pattern which would be absent in a rapid non-equilibrium transition 
with strong supercooling.  Since this issue has already received considerable 
attention in the literature~\cite{hydro}, we refrain from a detailed 
discussion here.

Further, as discussed in~\cite{Scavenius:1999zc}, if the chiral condensate 
is ``trapped'' in the false minimum and supercooling persists close to the 
spinodal instability, the subsequent realignment in the chiral field in the 
direction of the true vacuum can generate strong coherent pseudoscalar 
fields, $\vec{\pi}$, which will eventually decay into a ``beam'' of coherent 
pions.  The experimental situation will then be similar to that following 
an ``instantaneous quench'' of the restored symmetry phase~\cite{bja}.

Consider the behavior of the correlation length of the $\Phi$-field.
It is obvious from Fig.~\ref{potfig} that the curvatures of the effective 
potential at the minima $\Phi_f$ and $\Phi_t$ are rather different for 
$T<T_c$. Thus, the effective mass, $m_{\rm eff}^2={\rm d}^2 V_{\rm eff}/ 
{\rm d}\Phi^2$, and the correlation length, $l=1/m_{\rm eff}$, of the field 
will also be different. In particular, if the rapidly expanding system
supercools appreciably and if the order parameter is trapped in the
metastable state, one can expect a clear increase of $l$.
\begin{figure}[htp]
\centerline{\hbox{\epsfig{figure=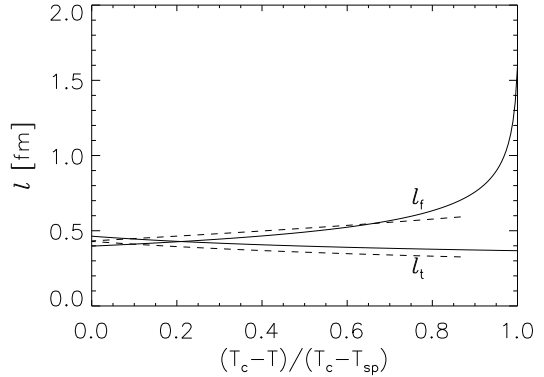,width=8cm}}}
\caption{Correlation length of the chiral order parameter in the metastable
state ($l_f$) and in the global minimum corresponding to an
equilibrium transition ($l_t$). Solid lines correspond to the exact results,
while dashed lines were obtained within the thin-wall approximation.}
\label{FigCorLength}
\end{figure} 
In Fig.~\ref{FigCorLength} we show $l$ at $\Phi_f$ and at $\Phi_t$ as
a function of the degree of supercooling.  The correlation length at 
$\Phi_f$ increases as the metastable minimum disappears.  By contrast, 
there is a smooth decrease in the correlation length at the true minimum 
$\Phi_t$ (which would be the expectation value of $\Phi$ if the transition 
proceeded in equilibrium).  In the present model $l_f$ can exceed $l_t$ by
as much as a factor of $2$--$3$, depending on the degree of supercooling.
The precise values of $l_f$ and $l_t$ depend on the specific effective model 
adopted, but the qualitative observation that $l_f$ increases dramatically 
as the system approaches the spinodal instability is general.  

At the critical temperature, the thin-wall approximation leads to only 
one correlation length, $l_{\rm tw} = \xi /2$, where $\xi$ is the 
thickness of the critical bubble as given by Eq.\ (3.12).  This is because 
the thin-wall approximation assumes both a small degree of supercooling 
and a small latent heat, i.e.\ a weak transition.  Thus, the curvatures at 
$\Phi_f$ and at $\Phi_t$ coincide to leading order in $\xi/R_c$.  To 
leading order in $\xi / R_c$ and for $T_{\rm sp} < T < T_c$, we find 
\be
W''(\phi_f) - W''(\phi_t) = - 6\left| j \right| /a ~.
\ee
Correlation lengths in the thin-wall approximation are shown in
Fig.~\ref{FigCorLength} by the dashed lines. The splitting of $l_f$ and $l_t$
is described reasonably well (at least qualitatively) except 
close to the spinodal, where this approximation is most unreliable.

We note that
the horizon, $1/H$, in expanding systems provides an upper bound on the correlation 
length; $l_f$ cannot increase beyond the horizon. Thus, observable 
consequences can be expected only if the expansion rate diminishes near  
the spinodal.  On the other hand, correlation lengths cannot become large 
if expansion rates are too small, e.g.\ $H < 0.1$~fm$^{-1}$, since the 
dynamics will no longer support strong supercooling.

As we have argued above, nucleation is not effective for small supercooling 
because $F_b / T$ is large.  The decay rate is then likely to be much smaller 
than the expansion time, and the system is likely to remain in the false 
vacuum.  On the other hand, the approximations of homogeneous nucleation 
theory are likely to break down near the spinodal where the decay rate of 
the false vacuum becomes large.  There, the condition $R_c \gg l_f$, 
inherent to nucleation theory, is no longer valid.  If $R_c \approx l_f$, 
it makes little sense to think of a bubble as a coherent superposition of 
random thermal fluctuations of wavelength $l_f$ in hydrostatic equilibrium 
with the surrounding symmetric phase.  It may rather be more appropriate to 
treat fluctuations on the scale $l_f$ as random thermal noise to be 
introduced into the classical equations of motion for the order parameter 
through, for example, a Langevin equation~\cite{langevin}.
In short, our results suggest 
that homogeneous nucleation theory may never provide a suitable modelling 
of the chiral phase transition in particle collisions.


\section{Summary and Outlook}

We have considered homogeneous bubble nucleation in a first-order chiral 
phase transition within an effective field theory approach to low-energy QCD, 
i.e.\ a linear $\sigma$-model coupled to quarks.  Integrating out the quarks 
leads to an effective potential for the soft modes of the chiral fields 
(i.e.\ the order parameter) with two competing minima for $T=T_c$, 
corresponding to a first-order phase transition.  The approximation of this 
effective potential by a familiar Landau-Ginzburg form allowed us to obtain 
analytical results for critical bubble profiles and decay rates using the 
thin-wall approximation.  Comparisons with exact numerical solutions, which 
do not emplay the thin-wall approximation, suggest that the thin-wall 
approximation is not adequate for the description of high-energy hadronic 
collisions.  Its use results in a significant underestimation of the size 
of critical bubbles.  The effects of this error are then amplified in the 
calculation of the decay rate.  Furthermore, the thin-wall approximation
can not be used to obtain the correlation length in the metastable state
close to the spinodal.

Our results show a rapid decrease with temperature 
in the free energy of critical bubbles 
from infinity at the critical temperature to zero at the 
spinodal.  Given the exponential dependence of $\Gamma$ on $F_b / T$, this 
fact provides some justification for our crude treatment of the prefactor 
in $\Gamma$.  Furthermore, it suggests that the metastable phase is likely to 
survive at temperatures such that $F_b \gg T$ is independent of the 
details of either the decay process or the expansion.  Even the weak 
first-order transition considered here suggests the viability of the 
restored symmetry phase for $(T_c - T)/(T_c - T_{\rm sp}) < 2/3$.  As 
we have suggested, stronger transitions will allow the metastable state to 
survive even closer to the spinodal instability.  The fate of the 
metastable state when $F_b < T$ is a more delicate issue and requires 
the comparison of the spatial/temporal scale, $\Gamma^{1/4}$, for critical 
thermal fluctuations to the typical expansion rate, $H$, expected in
high-energy collisions.  The small values of $\Gamma^{1/4}/H$ found in 
our calculations suggest that the phase transition is likely to proceed 
through spinodal decomposition rather than bubble nucleation for some 
fraction of all events (or some fraction of all rapidity bins).  Our 
results certainly indicate that $\Gamma^{1/4}$ is not materially larger 
than $H$.  This suggests the familiar idealized Maxwell construction of 
equilibrium thermodynamics is not appropriate for the description of 
phase transitions in high-energy heavy ion collisions. 
In fact, the relevance of nucleation theory for a possibly first-order
chiral phase transition in high-energy collisions
appears questionable. The conditions
$F_b < T$ and $l_f<R_c$ leave at best only a small window 
of temperatures where nucleation might occur.

Although current knowledge is insufficient to identify the order of the 
QCD phase transition in hadronic collisions with certainty, it is 
still important to search for experimental indications of supercooling.  
In addition to more familiar indicators, we have shown that the correlation 
length in the supercooled metastable state can be substantially larger than 
that expected in an equilibrium transition if the degree of supercooling is 
large.  Whether this can lead to observable consequences depends sensitively 
on the expansion rate of the system, $H$.  Small $H$ will not support 
deep supercooling, and the relevant correlation length will remain close 
to $l_t$, which is small. On the other hand, large $H$ will set causality
limits on the size of the correlation length at the false minimum, which 
cannot exceed the horizon $1/H$.  It remains to be seen if there is a
``window'' of energy or projectile/target combination for which the 
predicted increase in the correlation length can be observed. This 
speculation requires more detailed numerical studies of hydrodynamic expansion
coupled to dynamical bubble nucleation (with realistic initial conditions)
before definite predictions can be made.

The experimental observation of supercooling effects and spinodal 
decomposition would also be important as a matter of principle.  Ideally, 
signatures of phase transitions should be order parameter related and 
should reveal properties of the equilibrium phase diagram.  This is not 
the case for many of the signatures proposed for heavy ion collisions such 
as $J/\Psi$ suppression and strangeness production.  The interpretation 
of these signatures necessarily relies on theoretical estimates of 
reaction rates, and their connection to the thermodynamics of QCD is often 
far from clear.  While the spinodal instability is not part of the equilibrium 
phase diagram, it is very close.  Familiar experiments in condensed 
matter physics on a wealth of hysteresis phenomena (i.e.\ the analogue of 
superheating and supercooling) make it clear that spinodal instabilities 
can be studied on ``macroscopic'' time scales.  The spinodal instability, 
if found, would probably represent the most direct information which 
relativistic heavy ion collisions can provide regarding the QCD phase 
transition.  

Not surprisingly, our results indicate that the time-scales for
nucleation and expansion are not as well separated in high-energy
hadronic/heavy-ion collisions as they are, for example, in the cosmological QCD
phase transition.  The assumption of a ``well-mixed'' phase, commonly 
employed in hydrodynamic calculations~\cite{hydro} and the opposite 
scenario of perfect capture of the order parameter in the
metastable minimum~\cite{Scavenius:1999zc} with subsequent spinodal
decomposition, must be regarded as rough qualitative pictures.
Furthermore, the possibility of significant supercooling indicated by 
our results suggests the necessity of including the effects of 
non-spherical bubble configurations as well.  Bubble growth and bubble 
percolation in particular are likely to play an important role in completing 
the transition. If the system evolves by strong supercooling followed by a 
rapid phase transition (due either to spinodal decomposition or nucleation
of many small nonspherical bubbles which percolate), the constituent quarks
will experience strong reheating due to the fact that they become essentially
nonrelativistic. It would be interesting to seek experimental signatures 
for such ``reheating'', which must be present if the quarks acquire mass 
instantaneously.

In our view, a realistic and quantitative discussion of those points
requires numerical studies of the dynamical evolution of quarks (which
constitute the heat bath) coupled to the chiral field.  While initial 
investigations along these lines have been performed,
e.g. Refs.\ \cite{Scavenius:1999zc,CsM95}, the effects of dynamical
bubble nucleation have not yet been considered.  Moreover, the use of 
oversimplified geometries did not allow for realistic velocity and
temperature/density gradients, non-spherical bubble configurations, or bubble
percolation. Work on these questions is in progress and will be reported
elsewhere.


\acknowledgements
We thank A. Kusenko, D. Kharzeev, L. McLerran, I. Mishustin, R. Pisarski, and
R. Venugopalan for many fruitful comments and discussions. 
A.D. acknowledges support from the DOE Research Grant, Contract No.\
DE-FG-02-93ER-40764. E.S.F. is partially supported by the U. S. 
Department of Energy under Contract No.\ DE-AC02-98CH10886 and by 
CNPq (Brazil) through a post-doctoral fellowship.  J.T.L. thanks 
BNL's Nuclear Theory group for their generous 
support and hospitality during the
completion of this work.  J.T.L. also acknowledges the support of the
Director, Office of Energy Research, Division of Nuclear 
Physics of the Office of 
High Energy and Nuclear Physics of the U.S.\ Department of 
Energy under Contract No.\ DE-AC02-98CH10886.


\end{document}